# การศึกษาคุณภาพของบริการของโครงข่าย 5G ภาคสนาม: กรณีสถานีรถไฟฟ้าบีทีเอส
## A Field Trial Study of 5G Quality of Service: A Case of BTS Skytrain Station

เทอดพงษ์ แดงสี

พนา อังกาบ

พงษ์พิสิฐ วุฒิดิษฐโชติ

**บทคัดย่อ**

บทความนี้มีวัตถุประสงค์เพื่อศึกษาคุณภาพการให้บริการโครงข่าย 5G ของผู้ให้บริการหลัก 2 ราย โดยใช้วิธีทดสอบภาคสนามภายในบริเวณสถานีรถไฟฟ้ารถไฟฟ้าบีทีเอส 60 สถานี สำหรับผลการศึกษาคุณภาพการให้บริการซึ่งได้จากการทดสอบด้วย 4 แอปพลิเคชันบนโทรศัพท์เคลื่อนที่ 5G ผ่านโครงข่าย 5G ทั้ง 2 เครือข่าย แสดงให้เห็นว่าประสิทธิภาพของ 2 เครือข่ายมีความแตกต่างกัน เมื่อคำนวณหาค่าเฉลี่ยต่างๆ พบว่า มีค่าความเร็วในการดาวน์โหลดข้อมูลเท่ากับ 240.3 เมกะบิตต่อวินาที และมีค่าความเร็วในการอัปโหลดข้อมูลเท่ากับ 87.3 เมกะบิตต่อวินาที นอกจากสองค่านี้แล้ว ยังพบว่ามีค่าเฉลี่ยของค่าเวลาแฝง ค่าเวลาจิตเตอร์ และค่าความสูญเสียจากผู้ให้บริการทั้ง 2 ราย เท่ากับ 19 มิลลิวินาที 8 มิลลิวินาที และร้อยละ 0.299 ตามลำดับ เมื่อเปรียบเทียบค่าเฉลี่ยความเร็วจากผลการศึกษานี้กับผลการศึกษาในช่วงเวลาใกล้เคียงกันในภูมิภาคเอเชียแปซิฟิกที่เคยแสดงไว้ในรายงานฉบับหนึ่ง พบว่า ความเร็วในการดาวน์โหลดอยู่ในอันดับที่ 4 ส่วนความเร็วในการอัปโหลดอยู่ในอันดับที่ 1 อย่างไรก็ตาม พื้นที่ทดสอบในการศึกษานี้ครอบคลุมเฉพาะบริเวณสถานีรถไฟฟ้าบีทีเอสเท่านั้น ในอนาคตควรมีการขยายขอบเขตศึกษาให้ครอบคลุมพื้นที่อื่น ๆ ของกรุงเทพมหานครและปริมณฑล ตลอดจนพื้นที่ต่างจังหวัดอื่น ๆ ด้วย และการทดสอบดังกล่าวควรดำเนินการโดยองค์กรที่มีความน่าเชื่อถือ

**คำสำคัญ:** เทคโนโลยี 5G โทรศัพท์เคลื่อนที่ คุณภาพของบริการ ความเร็วในการดาวน์โหลด ความเร็วในการอัปโหลด

**Abstract**

This article aims to present the results from a study of the quality of service of 5G networks provided by two major 5G network providers, using a field trial approach within 60 BTS SkyTrain stations. The quality of service results obtained from the tests using 4 applications on a 5G mobile phone via two 5G networks showed that the performance of 5G networks provided by both operators are different. When calculated for the average values, it was found both operators had the download speed of 240.3 Mbps and the upload speed of 87.3 Mbps. Besides these values, it was also found the average latency, the average jitter and the average loss from both major operators that are 19 ms, 8 ms and 0.299 % respectively. Comparing the average values derived from the results of this study with the results of another study in the Asia-Pacific region that was undertaken during the similar period of time, it was found that the Bangkok download speed is ranked in the 4th place, whereas the upload speed is ranked in the 1st place. However, it is noted that this study only covered the area of BTS SkyTrain stations. In the future, the study should be extended to cover other parts of the Bangkok metropolitan area, and other provinces. In addition, the study should be conducted by a credible organization.

# 1. บทนำ

## 1.1 ความเป็นมาและความสำคัญ

เทคโนโลยีโทรคมนาคมเคลื่อนที่ยุคที่ 5 (The Fifth Generation of mobile communications technology: 5G) หรือเทคโนโลยี 5G เป็นมาตรฐานโทรคมนาคมที่ได้รับการพัฒนาหลักการและส่งเสริมโดยสหภาพโทรคมนาคมระหว่างประเทศ (International Telecommunication Union: ITU) ซึ่งเป็นหน่วยงานที่อยู่ภายใต้องค์การสหประชาชาติ ทั้งนี้ ITU ได้จัดทำข้อกำหนดสำหรับ 5G ในรูปแบบของมาตรฐาน IMT-2020 (เทอดพงษ์ แดงสี และ พิสิฐ พรพงศ์เตชวาณิช, 2562, น. 167) โดยได้เริ่มมีการส่งเสริมและพัฒนาหลักการดังกล่าวตั้งแต่ปี 2014 (ETSI, n.d.)

ในทางทฤษฎี เทคโนโลยี 5G มีคุณสมบัติหลายด้านที่ดีกว่าเทคโนโลยี 4G เดิม ซึ่งอธิบายได้พอสังเขปดังนี้ (เทอดพงษ์ แดงสี และ พิสิฐ พรพงศ์เตชวาณิช, 2562, น. 172; ETSI, n.d.)

1) สามารถรับส่งข้อมูลสูงสุดไม่น้อยกว่า 20 Gbps

2) สามารถสร้างประสบการณ์ที่ดีให้กับผู้ใช้งานด้วยอัตราการรับส่งข้อมูลเฉลี่ย ไม่น้อยกว่า 100 Mbps

3) สามารถใช้คลื่นความถี่ได้อย่างมีประสิทธิภาพ โดยเพิ่มขึ้นไม่น้อยกว่า 3 เท่า เมื่อเทียบกับ 4G

4) สามารถใช้งานบนยานพาหนะที่เคลื่อนที่ด้วยความเร็วสูงถึง 500 กิโลเมตรต่อชั่วโมง

5) สามารถลดค่าประวิงเวลาภายในเครือข่ายลงเหลือไม่เกิน 1 มิลลิวินาที

6) สามารถรองรับการเชื่อมต่อของอุปกรณ์ต่าง ๆ ได้พร้อมกันไม่น้อยกว่า 1 ล้านตัวต่อตารางกิโลเมตร

7) สามารถเพิ่มประสิทธิภาพด้านการใช้พลังงานได้ถึง 100 เท่า

8) รองรับการจราจรของข้อมูลในพื้นที่ 1 ตารางเมตรได้ไม่น้อยกว่า 10 กิโลบิตต่อวินาที

อย่างไรก็ตาม แม้ว่าในทางทฤษฎี เทคโนโลยี 5G จะมีคุณสมบัติต่าง ๆ ที่ดีเยี่ยม แต่ในทางปฏิบัติผู้ใช้งานยังไม่สามารถใช้งานคุณสมบัติเหล่านั้นได้อย่างเต็มประสิทธิภาพ ยกตัวอย่างเช่น ประสิทธิภาพในการรับส่งข้อมูล มีรายงานจากบริษัท Opensignal ระบุว่า ค่าเฉลี่ยของความเร็วในการดาวน์โหลดข้อมูลสูงที่สุด เฉลี่ย 415.6 เมกะบิตต่อวินาที ณ เมืองจอนจู สาธารณรัฐเกาหลี (เกาหลีใต้) ในขณะที่ค่าเฉลี่ยของความเร็วในการดาวน์โหลดข้อมูลของทั้งประเทศระบุว่า อันดับที่หนึ่งคือ สาธารณรัฐเกาหลี 380.5 เมกะบิตต่อวินาที อันดับที่สองและอันดับที่สามคือ สาธารณรัฐจีน (ไต้หวัน) 353.3 5 เมกะบิตต่อวินาที และเครือรัฐออสเตรเลีย 242.1 5 เมกะบิตต่อวินาที (Fogg, 2021a) ส่วนค่าเฉลี่ยของความเร็วในการอัปโหลดข้อมูลของทั้งประเทศระบุว่า อันดับที่หนึ่งคือสาธารณรัฐจีน (ไต้หวัน) 51.8 เมกะบิตต่อวินาที และอันดับที่สองคือ สาธารณรัฐเกาหลี 30.6 เมกะบิตต่อวินาที (Fogg, 2021a)

สำหรับประเทศไทย รายงานของ Opensignal อีกฉบับได้ระบุไว้ว่า มีความเร็วในการดาวน์โหลดข้อมูล (Download speed) ของโครงข่าย 5G ในประเทศไทยเฉลี่ย 122.5 เมกะบิตต่อวินาที และมีความเร็วในการอัปโหลดข้อมูล (Upload speed) เฉลี่ย 26.4 เมกะบิตต่อวินาที (Fogg, 2021b) ซึ่งทั้งสองค่าไม่สอดคล้องกับค่าเฉลี่ยที่วัดได้จากการศึกษาในระยะนำร่องของ Daengsi et al. (2021) ในพื้นที่หนึ่งในเขตกรุงเทพมหานครที่รายงานว่า มีความเร็วในการดาวน์โหลดข้อมูลของโครงข่าย 5G ในประเทศไทยเฉลี่ย 299.2 เมกะบิตต่อวินาที และมีความเร็วในการอัปโหลดข้อมูลเฉลี่ย 64.4 เมกะบิตต่อวินาที ทั้งนี้ Daengsi & Wuttidittachotti (2020) เคยตั้งข้อสังเกตถึงความน่าเชื่อถือของที่มาของข้อมูลที่ใช้ในการคำนวณหาค่าเฉลี่ยของความเร็วในการดาวน์โหลดข้อมูลจากรายงานของ Opensignal ที่เคยรายงานเกี่ยวกับความเร็วในการรับส่งข้อมูลของโครงข่าย 4G ในประเทศไทย

ดังนั้น เพื่อยืนยันผลการศึกษาในระยะนำร่องของ Daengsi et al. (2021) ที่มีผลแย้งกับรายงานการศึกษาของ Opensignal (Fogg, 2021b) คณะผู้วิจัยจึงได้ทำการศึกษาและวิเคราะห์ข้อมูลเพิ่มเติมจากงานวิจัยก่อนหน้านี้ (Daengsi et al., 2021) เพื่อนำเสนอผลการศึกษาที่มีหลักฐานเชิงประจักษ์และเพื่อพิสูจน์ให้เห็นถึงความเจริญก้าวหน้าของเทคโนโลยี 5G ที่กำลังมีการขยายโครงข่ายออกไปให้ครอบคลุมพื้นที่จังหวัดต่าง ๆ และคาดว่าจะกลายเป็นโครงสร้างพื้นฐานที่มีความสำคัญต่อสังคมและเศรษฐกิจของประเทศไทยในอนาคตอันใกล้

**1.2 การเปลี่ยนผ่านจาก 4G สู่ 5G**

หนึ่งในสาเหตุที่ประสิทธิภาพของเทคโนโลยี 5G ในปัจจุบันยังห่างไกลจากประสิทธิภาพที่ ITU กำหนด นอกจากจะเกิดจากข้อจำกัดต่าง ๆ คือ ข้อจำกัดจากการที่ผู้ให้บริการที่ยังไม่สามารถนำความถี่ย่านไฮแบนด์ (High-band) หรือมิลลิมีเตอร์เวฟ (Millimeter wave: mmWave) เช่น ความถี่ย่าน 26 กิกะเฮิรตซ์ มาใช้ได้ ณ เวลานี้ (O'Donnell, 2019) เนื่องจากจะต้องลงทุนในการวางระบบต่าง ๆ ใหม่ด้วยงบประมาณมูลค่ามหาศาล ดังนั้น ในระยะเริ่มต้นนี้ ผู้ให้บริการโครงข่าย 5G จึงใช้วิธีการนำโครงข่าย 4G ที่มีอยู่แล้วมาประยุกต์เข้ากับ

เทคโนโลยีสำหรับโครงข่าย 5G เพื่อให้สามารถให้บริการ 5G ได้ โดยเรียกว่า 5G แบบ Non-Standalone (5G NSA) ดังภาพที่ 1 (Son, 2019)

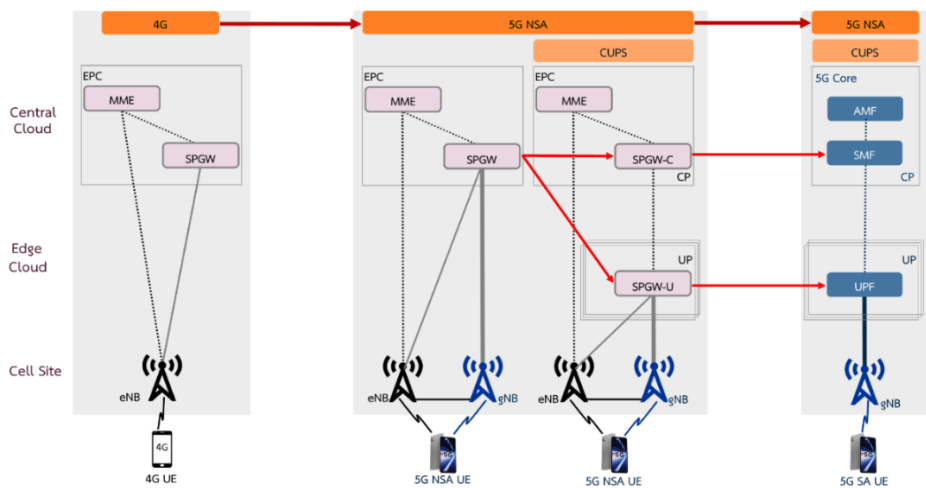

**ภาพที่ 1** การเปลี่ยนผ่านจาก 4G เป็น 5G

ที่มา: Son (2019)

อย่างไรก็ตาม ในอนาคตที่ผู้ให้บริการโครงข่าย 5G สามารถวางโครงข่าย 5G ได้เสร็จสมบูรณ์ ผู้ใช้งานก็จะได้รับประสบการณ์จากประสิทธิภาพที่เพิ่มขึ้นของ 5G แบบ Standalone (5G SA) ทั้งนี้ เครื่องโทรศัพท์ของผู้ใช้งานก็จะต้องรองรับ 5G SA ได้ด้วย (Gimme, 2020)

**2. การทบทวนวรรณกรรม**

นอกจากการศึกษารายงานของ Opensignal แล้ว (Fogg, 2021a, 2021b) คณะผู้วิจัยได้ทำการศึกษางานวิจัยภาคสนามอื่น ๆ ที่น่าสนใจเพื่อศึกษาถึงระเบียบวิธีวิจัย อย่างไรก็ตาม แม้ว่าจะพบงานวิจัยที่ศึกษากับประสิทธิภาพของโครงข่ายอื่น เช่น 3G (Chimmanee et al., 2015; วรวุฒิ ศาลางาม, 2015; พงศ์ปณต สุขปาน, 2015) แต่ในบทความนี้ทำการนำเสนอเฉพาะบทความที่มีการทดสอบประสิทธิภาพของ 5G ภาคสนามหรือบทความที่มีประเด็นใกล้เคียงกับบทความนี้เท่านั้น ดังแสดงในตารางที่ 1 ซึ่งจะเห็นได้ว่า มีการทดสอบประสิทธิภาพของโครงข่าย 5G ในภาคสนามด้วยการวัดหรือศึกษาพารามิเตอร์ของคุณภาพของบริการในหลายประเทศ และพบว่าหากไม่นับรวมบทความของ Daengsi et al. (2021) ที่เคยได้มีการศึกษาบริเวณสถานีรถไฟฟ้าบีทีเอสไปบางส่วนแล้ว ก็มีงานวิจัยที่มีประเด็นน่าสนใจอยู่ 2 บทความ ได้แก่ Okano et al. (2020) ที่มีการทดสอบสัญญาณ 28 กิกะเฮิรตซ์ GHz ของโครงข่าย 5G บริเวณชานชาลาสถานีรถไฟฟ้าที่ญี่ปุ่น ซึ่งคล้ายกับ

การศึกษาในบทความนี้ และ Zhao et al., (2020) ที่มีการทดสอบภาคสนามกับ 5G แบบ NSA ซึ่งเป็นการให้บริการโครงข่าย 5G ในระยะเปลี่ยนผ่านจาก 4G เป็น 5G โดยที่ยังอาศัยบางส่วนของโครงข่าย 4G เพื่อให้บริการ

**ตารางที่ 1** งานวิจัยที่เกี่ยวข้อง

| อ้างอิง | ค่าคุณภาพของบริการที่เกี่ยวข้อง | | | | | | พารามิเตอร์อื่น | โครงข่าย | | ลักษณะการทดสอบ | | เครื่องมือ | ประเทศ / หมายเหตุ |
|---|---|---|---|---|---|---|---|---|---|---|---|---|---|
| | Throughput | Download Speed | Upload Speed | Latency | Jitter | Loss | | 4G | 5G | อยู่กับที่ | เคลื่อนที่ | | |
| Huang et al. (2020) | ✓ | ✓ | - | ✓ | - | - | - | - | ✓ | ✓ | ✓ | เครื่องแม่ข่าย Microsoft Azure | 3 เมือง ในสหรัฐอเมริกา |
| Situmorang et al. (2019) | ✓ | - | - | - | - | - | - | - | ✓ | ✓ | | อุปกรณ์ยี่ห้อ Huawei | สาธารณรัฐอินโดนีเซีย |
| Liu et al. (2020) | - | ✓ | ✓ | - | - | - | - | - | ✓ | ไม่ระบุ | | Speedtest.cn | 105 เมือง ในสาธารณรัฐประชาชนจีน |
| Okumura et al. (2019) | ✓ | ✓ | ✓ | - | - | - | - | - | ✓ | ✓ | ✓ | ไม่ระบุ | ญี่ปุ่น |
| Okano et al. (2020) | ✓ | - | - | - | - | - | RSRP, SNR | - | ✓ | | | iPerf | |
| Abozariba et al. (2019) | ✓ | - | - | ✓ | - | - | RSSI | - | ✓ | ✓ | | เค้าโครงงานการเฝ้าระวัง | สหราชอาณาจักรบริเตนใหญ่และ |

| อ้างอิง | ค่าคุณภาพของบริการที่เกี่ยวข้อง | | | | | | พารามิเตอร์อื่น | โครงข่าย | | ลักษณะการทดสอบ | | เครื่องมือ | ประเทศ / หมายเหตุ |
|---|---|---|---|---|---|---|---|---|---|---|---|---|---|
| | Throughput | Download Speed | Upload Speed | Latency | Jitter | Loss | | 4G | 5G | อยู่กับที่ | เคลื่อนที่ | | |
| | | | | | | | | | | | | | ไอร์แลนด์เหนือ (พื้นที่ทุรกันดาร) |
| Tahir et al. (2019) | ✓ | ✓ | ✓ | ✓ | - | ✓ | - | ✓ | ✓ | | ✓ | สคริปต์ Python และคอมพิวเตอร์พีซีบนรถ | สาธารณรัฐฟินแลนด์ |
| Kutila et al. (2020) | - | - | - | ✓ | - | - | RSRP | - | ✓ | ✓ | | ไม่ระบุ | |
| Zhao et al. (2020) | ✓ | ✓ | ✓ | ✓ | - | - | MOS | ✓ | ✓ | ✓ | | ไม่ระบุ | สาธารณรัฐประชาชนจีน / 5G NSA |
| Daengsi & Wuttidittachotti, (2020) | | ✓ | ✓ | ✓ | ✓ | ✓ | - | ✓ | | ไม่ระบุ | | MIQ | ไทย |
| Daengsi et al. (2021) | - | ✓ | ✓ | ✓ | - | ✓ | - | - | ✓ | ✓ | | Speedtest | ไทย (สถานีรถไฟฟ้าบีทีเอส 11 สถานี) |

## 3. วิธีการศึกษา

### 3.1 การออกแบบการทดสอบ

การศึกษานี้ออกแบบให้ทดสอบในพื้นที่กรุงเทพมหานครเป็นพื้นที่หลักและเน้นศึกษาในพื้นที่ที่มีผู้ใช้โทรศัพท์เป็นจำนวนมาก คณะผู้วิจัยจึงได้ทำการเลือกศึกษาบริเวณสถานีรถไฟฟ้าบีทีเอส ซึ่งได้แนวคิดจากงานวิจัยก่อนหน้านี้ เช่น Okano et al. (2020) ที่มีการทดสอบภาคสนามบริเวณชานชาลาของสถานีรถไฟฟ้าในประเทศญี่ปุ่น และงานวิจัยที่เคยมีการดำเนินการในประเทศไทย เช่น Chimmanee et al. (2015) วรวุฒิ ศาลางาม (2015) และ พงศ์ปณต สุขปาน (2015) โดยการศึกษานี้ทำการทดสอบทั้งสายสุขุมวิท และสายสีลม เพราะวิ่งผ่านย่านธุรกิจและพื้นที่ที่มีประชากรหนาแน่น โดยเฉพาะสถานีหลัก 11 สถานี ได้แก่ ห้าแยกลาดพร้าว หมอชิต อนุสาวรีย์ชัยสมรภูมิ พญาไท สยาม ศาลาแดง ช่องนนทรี อโศก พร้อมพงษ์ อ่อนนุช และอุดมสุข เป็นสถานีที่มีผู้ใช้บริการเฉลี่ยวันละ 15,000–20,000 คนต่อวัน (Daengsi et al., 2021) โดยดำเนินการทดสอบอย่างเป็นระบบ (เฉลี่ยวันละ 10 สถานีโดยไม่ซ้ำกันยกเว้นสถานีสยามที่เป็นชุมทางที่เก็บข้อมูล 2 รอบ) ในระหว่างวันที่ 10-15 พฤษภาคม พ.ศ. 2564 ซึ่งครอบคลุม 6 วันของสัปดาห์ ในช่วงเวลา 8.00-18.00 น. ซึ่งครอบคลุมช่วงเวลาที่มีผู้ใช้บริการรถไฟฟ้ารวม 10 ชั่วโมง โดยเวลาในการทดสอบ ณ แต่ละสถานี ใช้รูปแบบการสุ่มที่เป็นอิสระต่อกัน ทั้งนี้ ข้อมูลที่สำคัญที่เกี่ยวข้องกับการออกแบบและการเลือกใช้เครื่องมือในการทดสอบได้แสดงไว้ในตารางที่ 2

อย่างไรก็ตาม ในการทดสอบแต่ละสถานีเพื่อทดสอบและเก็บข้อมูลคุณภาพของบริการ ผู้วิจัยได้ออกแบบให้มีการทดสอบสถานีละ 4 จุดๆ ละ 4 แอปพลิเคชัน (ชั้นจำหน่ายตั๋ว 3 จุด และชั้นชานชาลาที่อยู่ติดกับชั้นจำหน่ายตั๋ว 1 จุด ดังแสดงในภาพที่ 2) โดยทดสอบกับโครงข่าย 5G ของผู้ให้บริการหลัก 2 ราย ด้วยเครื่องโทรศัพท์ที่รองรับ 5G ได้ เมื่อทดสอบเสร็จแต่ละครั้ง ผู้ทดสอบมีการบันทึกภาพหน้าจอทุกครั้ง เพื่อนำกลับไปทำการบันทึกข้อมูลในภายหลัง

**ตารางที่ 2** ข้อมูลเกี่ยวกับวิธีการทดสอบและเครื่องมือ

| รายการ | รายละเอียด |
| --- | --- |
| พื้นที่ที่ทดสอบ | บริเวณสถานีรถไฟฟ้าบีทีเอส สายสุขุมวิท (คูคต จังหวัดปทุมธานี - เคหะฯ สมุทรปราการ จังหวัดสมุทรปราการ) และสายสีลม (สนามกีฬาแห่งชาติ-บางหว้า) รวม 60 สถานี |
| วัน-เวลาที่ทดสอบ | 10-15 พฤษภาคม พ.ศ. 2564 เวลา 8:00-18:00 น. |
| จำนวนโครงข่ายที่ทดสอบ | 2 โครงข่าย จาก 2 ผู้ให้บริการหลัก |
| เครื่องโทรศัพท์ | 1 เครื่อง ยี่ห้อหัวเหว่ย รุ่น P40 Pro |
| แอปพลิเคชันที่ใช้ทดสอบ | 4 แอปพลิเคชัน ได้แก่ nPerf, Opensignal, Speedtest และ Speed Master |
| ปริมาณข้อมูลที่เก็บ | 1,920 รายการ |

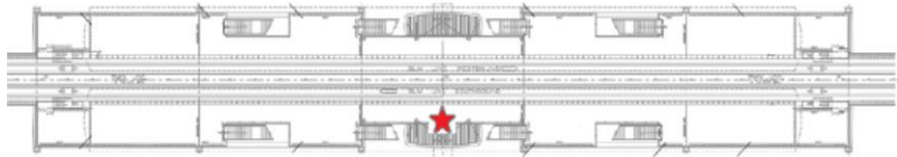

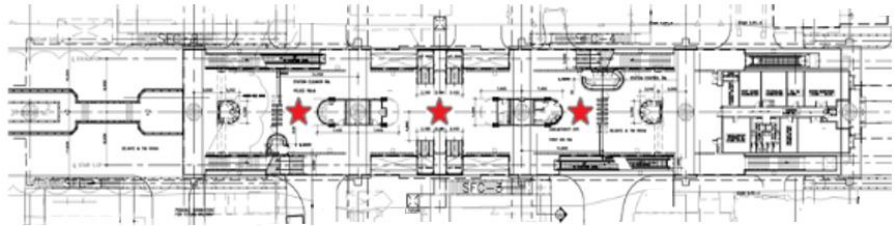

**ภาพที่ 2** มุมมองด้านบน (ก) จุดทดสอบบนชั้นชานชาลา (ข) จุดทดสอบบนชั้นจำหน่ายตั๋ว

**3.2 เครื่องมือที่ใช้ในการทดสอบภาคสนาม** ประกอบด้วยเครื่องมือหลักดังนี้

3.2.1 โทรศัพท์ยี่ห้อหัวเว่ย รุ่น P40 Pro ซึ่งรองรับได้ 2 ซิม (Subscriber Identity Module: SIM) และมีประสิทธิภาพสูงเมื่อเทียบกับหลายรุ่นของหลายยี่ห้อ ณ ช่วงเวลานั้น เพราะหากใช้รุ่นที่มีประสิทธิภาพต่ำ อาจทำให้ได้ผลการทดสอบที่ต่ำกว่าที่ควรจะเป็นเนื่องจากติดข้อจำกัดเรื่องประสิทธิภาพของหน่วยประมวลผลของเครื่องโทรศัพท์ที่ใช้เป็นเครื่องมือ โดยเครื่องโทรศัพท์ที่เลือกใช้ในการศึกษานี้ใช้ชิปเซต Kirin 990 5G ซึ่งเป็นแบบ Octa-core และใช้เทคโนโลยี 7nm และสามารถรองรับคลื่นความถี่ที่มีการอนุญาตให้ใช้บริการในโครงข่าย 5G ในประเทศไทยในปัจจุบันได้ (GSMArena, n.d.) นอกจากนี้ยังเป็นเครื่องที่ใช้ระบบปฏิบัติการ Android ของ Google ซึ่งเป็นระบบปฏิบัติการที่มีส่วนแบ่งการตลาดสูงที่สุดเมื่อเทียบกับระบบปฏิบัติการอื่น คือ มากกว่า 85% (โดยประมาณ) (O'Dea, 2021)

3.2.2 ซิมเลขหมายโทรศัพท์ของผู้ให้บริการหลัก 2 ราย (ในที่นี้จะเรียกว่า Oper1 และ Oper2 สำหรับผู้ให้บริการรายที่ 1 และ 2 ตามลำดับ) พร้อมแพ็กเกจที่รองรับบริการ 5G แบบไม่จำกัด ซึ่งเลือกจากผู้ให้บริการโครงข่ายโทรศัพท์เคลื่อนที่ในประเทศไทยซึ่งปัจจุบันมี 4 ราย ประกอบด้วยรัฐวิสาหกิจ 1 ราย และผู้ประกอบการที่เป็นบริษัทเอกชน 3 ราย โดย 3 รายนี้มียอดรวมผู้ลงทะเบียนประมาณ 42.8 ล้านเลขหมาย 31.2 ล้านเลขหมาย และ 19.1 ล้านเลขหมาย ตามลำดับ (กรุงเทพธุรกิจ, 2564)

3.2.3 แอปพลิเคชันทดสอบความเร็วของอินเทอร์เน็ต จำนวน 4 แอปพลิเคชัน ซึ่งเลือกจากรายชื่อแอปพลิเคชันที่มีผู้ใช้งานนิยมใช้งาน และส่วนใหญ่เป็นแอปพลิเคชันที่ผู้ใช้งานและผู้ให้บริการค่อนข้างคุ้นเคย ประกอบด้วย nPerf, Opensignal, Speedtest และ Speed Master ดังแสดงในตารางที่ 2 (Hindy, 2021; Newzoogle, n.d.)

## 4. ผลการศึกษา

จากการทดสอบและเก็บข้อมูลภาคสนามบริเวณสถานีรถไฟฟ้าบีทีเอส (แอปพลิเคชันละ 240 ครั้ง รวม 1,920 ครั้ง) ได้มีกระบวนการตรวจหาข้อมูลที่มีลักษณะเป็นค่าผิดปกติ (Outliers) แล้วตัดออก เช่น ข้อมูลชุดที่มีค่าความเร็วเฉลี่ยในการอัปโหลดสูงมากกว่า 1,200 เมกะบิตต่อวินาทีทั้งที่เป็นการเชื่อมต่อกับโครงข่าย 4G ไม่ใช่ 5G และมีเวลาแฝงมากกว่า 1,500 มิลลิวินาที เป็นต้น (ดังแสดงในภาพที่ 3) ก่อนที่จะทำการวิเคราะห์ด้วยสถิติแบบพรรณนาแล้วทำการรายงานผลค่าพารามิเตอร์ของคุณภาพของบริการ ที่ใช้ในการพิจารณาประสิทธิภาพของเครือข่ายซึ่งประกอบด้วย ค่าความเร็วในการดาวน์โหลดข้อมูล ค่าความเร็วในการอัปโหลดข้อมูล ค่าเวลาแฝง (Latency) (บางครั้งใช้ค่าประวิงเวลา (Delay) แทน) ค่าจิตเตอร์ (Jitter) และค่าสูญเสีย (Loss) ซึ่งเมื่อทำการหาค่าเฉลี่ยของค่าพารามิเตอร์ของคุณภาพของบริการของโครงข่าย 5G จากข้อมูลของ Oper1 จำนวน 916 รายการ และ Oper2 จำนวน 958 รายการ สามารถแสดงผลการศึกษาได้ดังภาพที่ 4-8

อย่างไรก็ตาม จากการศึกษานี้ผู้วิจัยพบประเด็นว่า ทุกแอปพลิเคชันสามารถวัดค่าความเร็วในการดาวน์โหลด (Download: DL) อัปโหลด (Upload: UL) และค่าเวลาแฝงได้ แต่บางแอปพลิเคชัน เช่น nPerf และ Opensignal วัดค่าความสูญเสียไม่ได้ แม้แต่แอปลิเคชัน Speed Master ที่แสดงข้อมูลว่าสามารถวัดค่าความสูญเสียได้ แต่ในทางปฏิบัติพบว่าแสดงค่าความสูญเสียเป็นศูนย์ทุกครั้งที่ทดสอบ โดยพบว่ามีเพียงแอปพลิเคชัน Speedtest เพียงแอปพลิเคชันเดียวจากทั้งหมด 4 แอปพลิเคชัน ที่สามารถวัดค่าความสูญเสียได้

จากภาพที่ 4-5 จะเห็นได้ว่า ผลการวัดได้จากต่างแอปพลิเคชันกัน ให้ค่าเฉลี่ยความเร็วในการดาวน์โหลดและอัปโหลดที่ต่างกัน อย่างไรก็ตามพบว่า ค่าโดยภาพรวม Oper1 มีค่าเฉลี่ยความเร็วในการดาวน์โหลดและอัปโหลดสูงกว่า Oper2 และเมื่อเฉลี่ยรวมทั้ง 4 แอปพลิเคชัน พบว่า Oper1 และ Oper2 มีค่าเฉลี่ยความเร็วในการดาวน์โหลดอยู่ที่ 314.4 เมกะบิตต่อวินาที และ 169.5 เมกะบิตต่อวินาที ตามลำดับ ในขณะที่ค่าเฉลี่ยรวมจากผู้ให้บริการโครงข่าย 5G ทั้งสองราย มีค่าเฉลี่ยความเร็วในการดาวน์โหลดอยู่ที่ 240.3 เมกะบิตต่อวินาที สำหรับค่าเฉลี่ยความเร็วในการอัปโหลดพบว่า Oper1 และ Oper2 อยู่ที่ 126.1 เมกะบิตต่อวินาที และ 50.2 เมกะบิตต่อวินาที ตามลำดับ ในขณะที่ค่าเฉลี่ยรวมจากผู้ให้บริการโครงข่าย 5G ทั้งสองราย มีค่าเฉลี่ยความเร็วในการดาวน์โหลดอยู่ที่ 87.3 เมกะบิตต่อวินาที

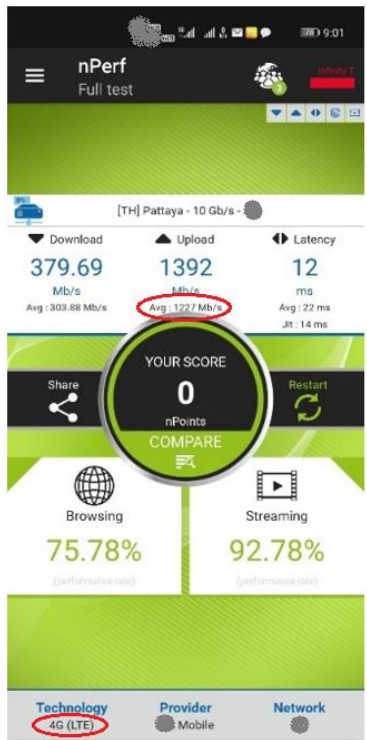 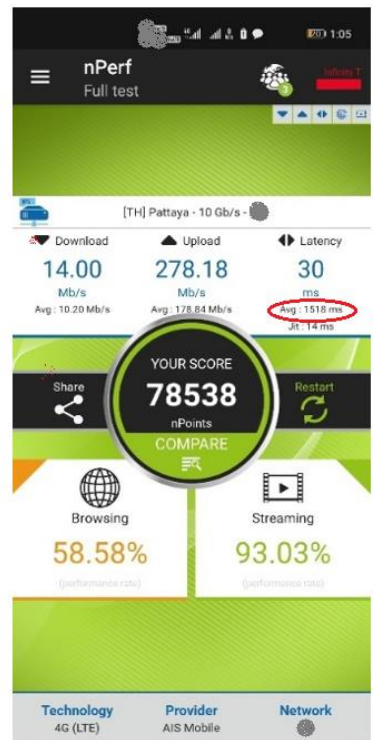

(ก)　　　　　　　　　　　　(ข)

**ภาพที่ 3** (ก) ความเร็วในการอัปโหลดสูงผิดปกติทั้งที่เชื่อมต่อโครงข่าย 4G และ (ข) ค่าเวลาแฝงเฉลี่ยสูงผิดปกติ

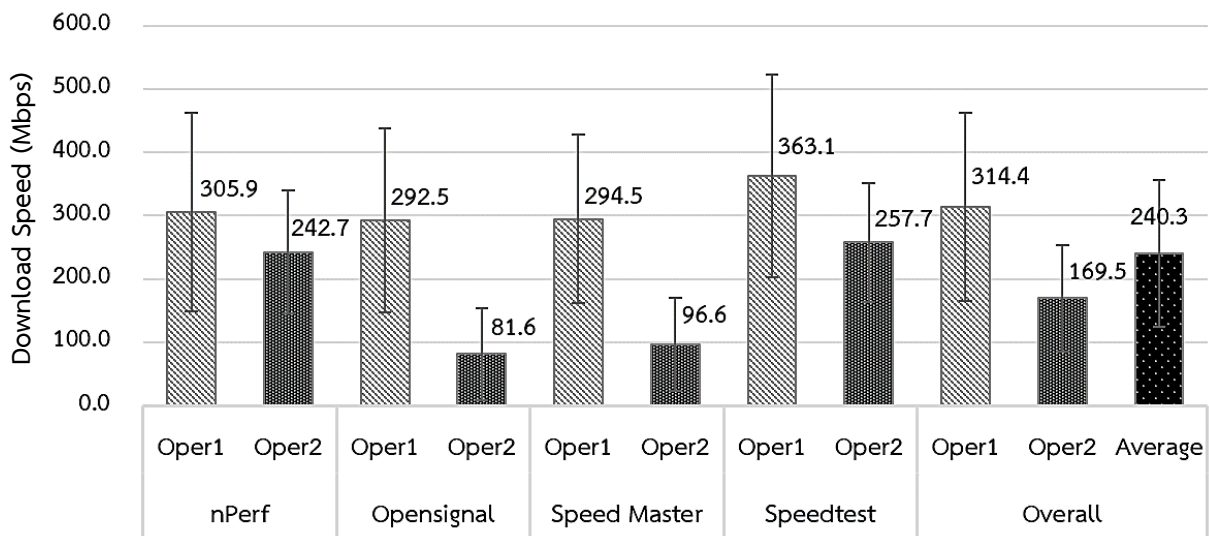

**ภาพที่ 4** ค่าเฉลี่ยความเร็วในการดาวน์โหลดข้อมูลผ่านโครงข่าย 5G จากการทดสอบบนสถานีรถไฟฟ้าบีทีเอส

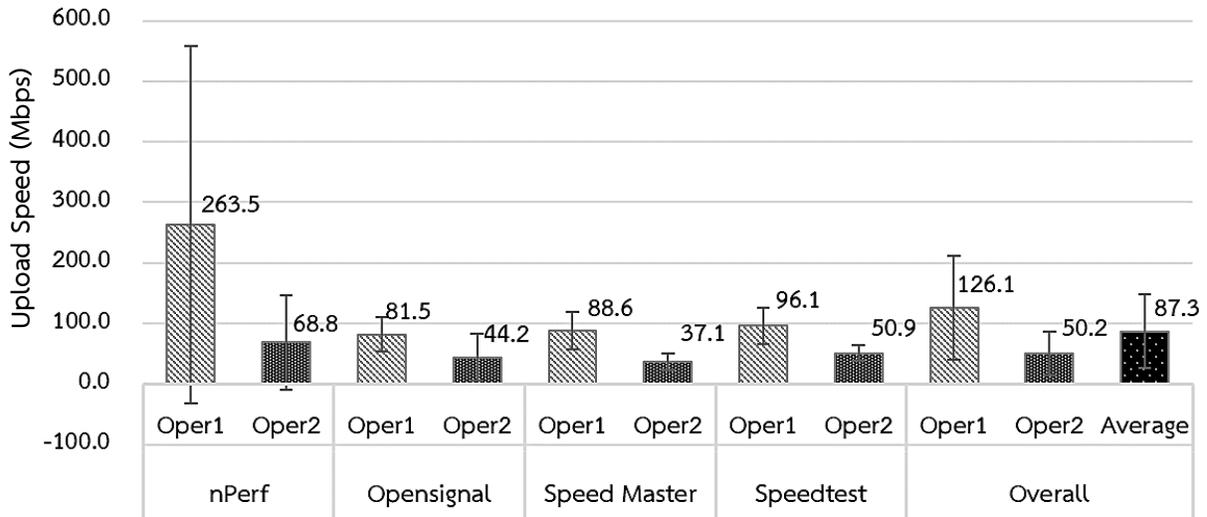

**ภาพที่ 5** ค่าเฉลี่ยความเร็วในการอัปโหลดข้อมูลผ่านโครงข่าย 5G จากการทดสอบบนสถานีรถไฟฟ้าบีทีเอส

สำหรับภาพที่ 6 เป็นภาพแสดงผลของค่าเวลาแฝง จากการทดสอบภาคสนาม ซึ่งจะเห็นได้ว่า ในภาพรวมค่าเวลาแฝงของ Oper1 มีค่าสูงกว่า Oper2 เล็กน้อย โดยที่ 3 ใน 4 แอปพลิเคชัน มีค่าเวลาแฝงอยู่ในช่วง 10-20 มิลลิวินาที แต่ค่าที่วัดด้วย Speed Master จาก Oper1 และ Oper2 มีค่าเฉลี่ยเท่ากับ 36.8 มิลลิวินาที และ 30.6 มิลลิวินาที ตามลำดับ ทำให้ค่าเฉลี่ยรวมจากทั้ง 4 แอปพลิเคชัน มีค่าเวลาแฝงอยู่ที่ 21.5 มิลลิวินาที และ 16.6 มิลลิวินาที สำหรับ Oper1 และ Oper2 ตามลำดับ ซึ่งเฉลี่ยรวมกันเท่ากับ 19 มิลลิวินาที

ส่วนภาพที่ 7 แสดงผลของค่าจิตเตอร์ ซึ่งเป็นค่าความแปรปรวนของค่าเวลาแฝงที่วัดได้จากการทดสอบภาคสนาม จึงมีลักษณะคล้ายกันกับค่าเวลาแฝง คือค่านี้ยิ่งน้อยยิ่งดี จากภาพนี้จะเห็นได้ว่า ในภาพรวมค่าจาก Oper1 มีค่าสูงกว่า Oper2 เล็กน้อย ยกเว้นค่าที่วัดได้จาก Speedtest ที่พบว่า Oper2 มีค่าสูงหรือแย่กว่าค่าจาก Oper1 เพียงเล็กน้อย โดยค่าเฉลี่ยรวมของจิตเตอร์จากทั้งสองโครงข่ายในการศึกษานี้มีค่าเท่ากับ 8 มิลลิวินาที

ในภาพที่ 8 ซึ่งเป็นภาพที่แสดงค่าความสูญเสียในการรับส่งข้อมูลผ่านโครงข่าย 5G ที่ได้จากการทดสอบบนสถานีรถไฟฟ้าบีทีเอส จะเห็นได้ว่า ค่านี้สามารถวัดได้ด้วยแอปพลิเคชันเดียวเท่านั้นคือ Speectest โดยพบว่า Oper1 มีค่าเฉลี่ยความสูญเสียอยู่ที่ร้อยละ 0.595 ซึ่งสูงกว่าค่าเฉลี่ยจากโครงข่ายของ Oper2 ที่มีค่าเพียงร้อยละ 0.003 อย่างไรก็ตาม เมื่อคำนวณค่าเฉลี่ยความสูญเสียจากทั้งสองโครงข่ายพบว่ามีค่าร้อยละ 0.299

นอกจากนี้ เมื่อพิจารณาจากข้อมูลที่ได้จากแอปพลิเคชัน nPerf ที่ตัดค่าผิดปกติออกแล้วพบว่าบริการจากโครงข่ายของ Oper1 และ Oper2 พบว่าร้อยละ 54.6 ของระบบที่ให้บริการภายใต้โครงข่ายของ Oper1 และร้อยละ 60.8 ของระบบที่ให้บริการภายใต้โครงข่ายของ Oper2 เป็นระบบ 5G NSA ดังแสดงในตารางที่ 3 นอกนั้นเป็นระบบ 4G หรือ LTE โดยไม่พบว่ามีการทดสอบครั้งใดได้รับบริการจากระบบ 5G SA เลย นอกจากนี้ยังพบความผิดปกติบางอย่าง เช่น ในการทดสอบด้วยแอปพลิเคชัน Speed Master มีการแสดงค่าความสูญเสียเป็นศูนย์ทุก

ครั้งที่มีการทดสอบ และในการทดสอบด้วยแอปพลิเคชัน Opensignal ผ่านเลขหมายของ Oper2 แต่แอปพลิเคชัน Opensignal แสดงข้อมูลโครงข่ายเป็น Oper1 ทุกครั้ง

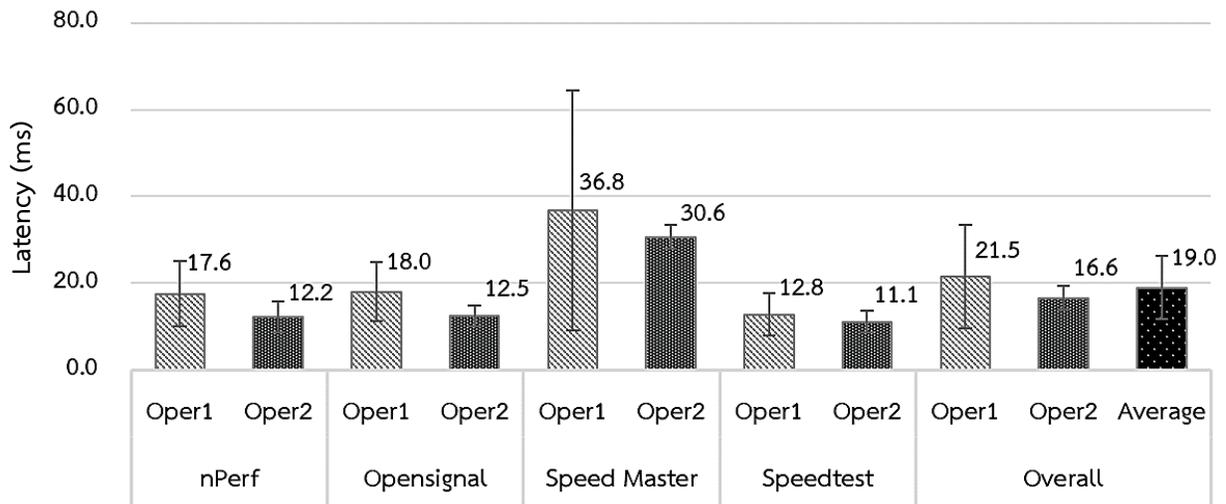

**ภาพที่ 6** ค่าเวลาแฝงเฉลี่ยในโครงข่าย 5G ที่ได้จากการทดสอบบนสถานีรถไฟฟ้าบีทีเอส

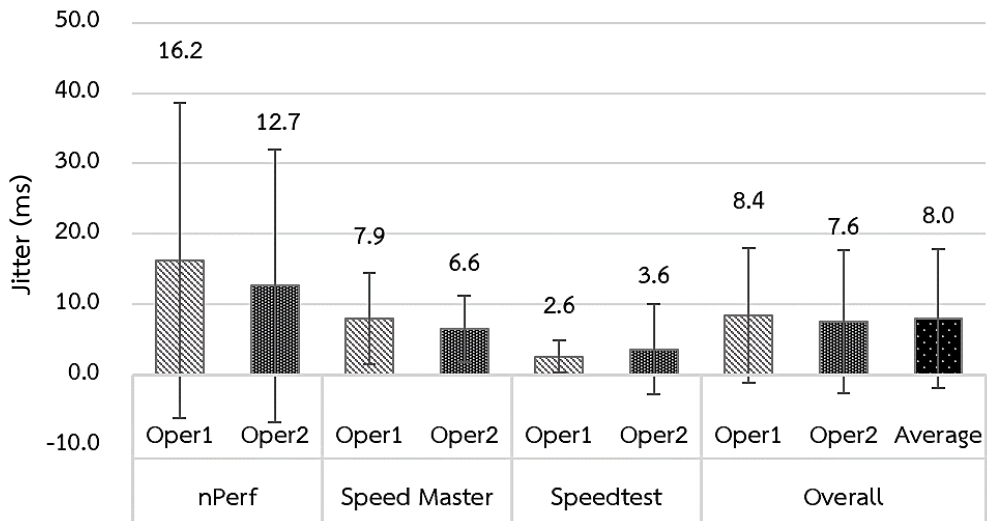

**ภาพที่ 7** ค่าเวลาจิตเตอร์เฉลี่ยในโครงข่าย 5G ที่ได้จากการทดสอบบนสถานีรถไฟฟ้าบีทีเอส

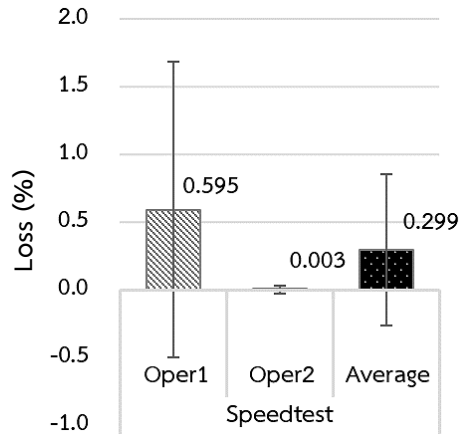

**ภาพที่ 8** ค่าความสูญเสียในการรับส่งข้อมูลผ่านโครงข่าย 5G ที่ได้จากการทดสอบบนสถานีรถไฟฟ้าบีทีเอส

**ตารางที่ 3** ข้อระบบที่ให้บริการในระหว่างการทดสอบภาคสนาม

| ผู้ให้บริการ | ระบบที่ให้บริการ | | | หมายเหตุ |
|---|---|---|---|---|
| | 5G SA | 5G NSA | 4G (LTE) | |
| Oper1 | 0 | 107 | 89 | 107/(107+89)*100 = 54.6 % |
| Oper2 | 0 | 141 | 91 | 141/(141+91)*100 = 60.8 % |

## 5. การอภิปรายผล

จากผลการทดสอบด้วย 4 แอปพลิเคชัน สามารถอภิปรายได้เป็นแต่ละประเด็นดังนี้

5.1 สำหรับค่าเฉลี่ยความเร็วในการดาวน์โหลด 240.3 เมกะบิตต่อวินาที ในเขตพื้นที่กรุงเทพมหานครและปริมณฑลบางส่วน จาก 4 แอปพลิเคชัน จะเห็นได้ว่า หากเทียบกับข้อมูลที่รายงานโดย Opensignal (Fogg, 2021b) ซึ่งใช้หลักการในการเก็บข้อมูลจากผู้ใช้งานในหลายประเทศทั่วโลกที่มีการติดตั้งแอปพลิเคชันของ Opensignal ในช่วงเวลาใกล้เคียงกันและทดสอบโดยใช้งาน แล้ว Opensignal ได้ทำการวิเคราะห์ข้อมูลที่ได้และคำนวณหาค่าต่าง ๆ เช่น ค่าเฉลี่ยความเร็วในการดาวน์โหลดและอัปโหลดในประเทศหรือเมืองต่าง ๆ (Opensignal, 2020) ค่าที่ได้จากการศึกษานี้สูงกว่าค่าเฉลี่ยความเร็วในการดาวน์โหลดในประเทศญี่ปุ่น (180.9 เมกะบิตต่อวินาที) โดยเป็นรองเฉพาะสาธารณรัฐเกาหลี (380.5 เมกะบิตต่อวินาที) สาธารณรัฐจีน (ไต้หวัน) (353.3 เมกะบิตต่อวินาที) และเครือรัฐออสเตรเลีย (242.1 เมกะบิตต่อวินาที) (ดังที่ผู้เขียนได้สรุปไว้ในตารางที่ 4) อย่างไรก็ตาม หากพิจารณาเฉพาะค่าเฉลี่ยจากแอปพลิเคชัน Opensignal ค่าความเร็วในการดาวน์โหลดจาก Oper1 และ Oper2 จะได้ค่าเฉลี่ยประมาณ (292.5+81.6)/2 ≈ 187.1 เมกะบิตต่อวินาที (ดูภาพที่ 4) ซึ่งก็ยังสูง

กว่าค่าเฉลี่ยของประเทศญี่ปุ่น แม้ว่าจะยังมีค่าน้อยกว่าสาธารณรัฐเกาหลี สาธารณรัฐจีน (ไต้หวัน) และเครือรัฐออสเตรเลียก็ตาม นอกจากนี้ เมื่อเปรียบเทียบกับผลการศึกษาค่าความเร็วในการดาวน์โหลดของโครงข่าย 4G ในประเทศไทยที่เคยมีการศึกษาเมื่อ พ.ศ. 2563 (Daengsi & Wuttidittachotti, 2020) ถือว่าเร็วขึ้นประมาณ 10 เท่า ซึ่งค่าเดิมจากโครงข่าย 4G ที่เคยมีการศึกษามีค่าเฉลี่ย 17.8 เมกะบิตต่อวินาที

    5.2 สำหรับค่าเฉลี่ยความเร็วในการอัปโหลด 87.3 เมกะบิตต่อวินาที ในเขตพื้นที่กรุงเทพมหานครจาก 4 แอปพลิเคชัน หากเทียบกับข้อมูลที่รายงานโดย Opensignal (Fogg, 2021b) จะเห็นได้ว่า เป็นค่าที่สูงที่สุดเทียบกับค่าเฉลี่ยความเร็วในการดาวน์โหลดของทุกประเทศในภูมิภาคเอเชียแปซิฟิก (ดังที่ผู้เขียนได้สรุปไว้ในตารางที่ 4) อย่างไรก็ตาม หากพิจารณาเฉพาะค่าเฉลี่ยจากแอปพลิเคชัน Opensignal ค่าความเร็วในการอัปโหลดจาก Oper1 และ Oper2 จะได้ค่าเฉลี่ยประมาณ (81.5+44.2)/2 ≈ 62.9 เมกะบิตต่อวินาที (ดูภาพที่ 5) ซึ่งก็ยังเป็นค่าที่สูงที่สุดเทียบกับค่าเฉลี่ยความเร็วในการดาวน์โหลดของทุกประเทศในภูมิภาคเอเชียแปซิฟิก เพราะในรายงานของ Opensignal ระบุว่าค่าเฉลี่ยความเร็วในการอัปโหลดสูงสุดเป็นค่าที่วัดได้จากสาธารณรัฐจีน (ไต้หวัน) โดยมีค่าเท่ากับ 51.8 เมกะบิตต่อวินาที นอกจากนี้ เมื่อเปรียบเทียบกับผลการศึกษาค่าความเร็วในการอัปโหลดของโครงข่าย 4G ในประเทศไทยที่เคยมีการศึกษาเมื่อ พ.ศ. 2563 (Daengsi & Wuttidittachotti, 2020) ถือว่าเร็วขึ้นประมาณ 4 เท่า ซึ่งค่าเดิมจากโครงข่าย 4G ที่เคยมีการศึกษามีค่าเฉลี่ย 14.6 เมกะบิตต่อวินาที

**ตารางที่** 4 เปรียบเทียบผลการศึกษานี้กับผลการศึกษาอื่น (Fogg, 2021b; Daengsi & Wuttidittachotti, 2020)

| อันดับ | การดาวน์โหลด | | การอัปโหลด | |
|---|---|---|---|---|
| | ประเทศ | ความเร็ว (เมกะบิต/วินาที) | ประเทศ | ความเร็ว (เมกะบิต/วินาที) |
| 1 | สาธารณรัฐเกาหลี | 380.5** | ผลการศึกษานี้ | **87.3*** |
| 2 | สาธารณรัฐจีน (ไต้หวัน) | 353.3** | สาธารณรัฐจีน (ไต้หวัน) | 51.8** |
| 3 | เครือรัฐออสเตรเลีย | 242.1** | สาธารณรัฐเกาหลี | 30.6** |
| 4 | ผลการศึกษานี้ | **240.3*** | ไทย | 26.4** |
| 5 | ญี่ปุ่น | 180.9** | เขตบริหารพิเศษฮ่องกงแห่งสาธารณรัฐประชาชนจีน | 21.5** |
| 6 | ไทย | 122.5** | สิงคโปร์ | 21.2** |

| 7 | ไทย (4G) | 17.8*** | ไทย (4G) | 14.6*** |

หมายเหตุ: \*  เป็นผลการศึกษานี้

\*\*  เป็นผลการศึกษาจาก Fogg (2021b)

\*\*\* เป็นผลการศึกษาจาก Daengsi & Wuttidittachotti, 2020

     5.3 จากตารางที่ 4 จะเห็นได้ว่า ค่าเฉลี่ยความเร็วในการดาวน์โหลดและอัปโหลดในการศึกษานี้มีค่า 240.3 เมกะบิตต่อวินาที และ 87.3 เมกะบิตต่อวินาที ตามลำดับ ซึ่งมีค่าสูงกว่าผลการศึกษาของ Fogg (2021b) ที่ค่าเฉลี่ยความเร็วในการดาวน์โหลดและอัปโหลดมีค่า 122.5 เมกะบิตต่อวินาที และ 26.4 เมกะบิตต่อวินาที ตามลำดับ อย่างเห็นได้ชัด ซึ่งอาจอธิบายได้ว่า การศึกษานี้อยู่ในบริเวณสถานีรถไฟฟ้าบีทีเอส ซึ่งอยู่ในพื้นที่กรุงเทพฯและปริมณฑล เป็นบริเวณที่มีประชาชนสัญจรผ่านไปมาในแต่ละสถานีเป็นจำนวนมาก จึงอนุมานได้ว่า ผู้ให้บริการ 5G ต่างเห็นความสำคัญจึงมีการวางโครงข่าย 5G ครอบคลุมสถานีรถไฟฟ้าบีทีเอส ค่าที่ได้จากการศึกษานี้จึงสูงกว่าของ Fogg (2021b) ที่เป็นการศึกษาจากข้อมูลแบบคราวด์ซอร์ส (Crowdsourced data) จากผู้ใช้บริการ 5G จากทั่วประเทศ ซึ่งบางจังหวัดอาจเพิ่งเริ่มติดตั้งโครงข่าย 5G ในช่วงเวลานั้น จึงอาจยังไม่สามารถให้บริการได้เต็มประสิทธิภาพ นอกจากนี้ การที่ Fogg (2021b) ศึกษาจาก Crowdsourced data ทำให้ไม่สามารถกำหนดได้ว่าให้ผู้ใช้บริการ 5G แต่ละคนใช้เครื่องโทรศัพท์รุ่นใดในการทดสอบ ซึ่งประสิทธิภาพของหน่วยประมวลผลกลางของเครื่องโทรศัพท์สามารถส่งผลกระทบต่อผลการทดสอบได้ แตกต่างจากการศึกษานี้ที่ใช้เครื่องโทรศัพท์รุ่นที่เป็นเรือธง (Flagship) ของยี่ห้อหัวเหว่ย ที่ทำตลาดในช่วงกลางปี พ.ศ. 2563 ถึงต้นปี พ.ศ. 2564 ซึ่งมีประสิทธิภาพสูง เนื่องจากทำงานด้วยหน่วยประมวลผลกลางที่มีประสิทธิภาพสูง นอกจากนี้ ในการที่การศึกษาของ Fogg (2021) ดำเนินการในระหว่างเดือนมีนาคม-พฤษภาคม 2564 ก็อาจเป็นอีกหนึ่งสาเหตุที่ทำให้ผลการศึกษาที่ได้มีความแตกต่างกัน หากต้นเดือนพฤษภาคมที่เป็นช่วงเวลาของการดำเนินการศึกษานี้ ผู้ให้บริการมีการพัฒนาระบบ 5G ที่ดีกว่า และมีการติดตั้งโครงข่าย 5G ที่ครอบคลุมพื้นที่มากกว่าช่วงเดือนมีนาคม-เมษายน 2564

     5.4 เมื่อเปรียบเทียบผลการศึกษานี้กับผลการศึกษาในระยะนำร่อง (Daengsi et al., 2021) พบว่า ค่าเฉลี่ยความเร็วในการดาวน์โหลดในบทความนี้ซึ่งเป็นค่าเฉลี่ยจาก 4 แอปพลิเคชันมีค่าเฉลี่ยที่ต่ำกว่า เนื่องจากผลการศึกษาในระยะนำร่องเป็นผลการทดสอบจากแอปพลิเคชัน Speedtest เพียงแอปพลิเคชันเดียว คือ Speedtest ซึ่งปกติให้ผลการวัดความเร็วในการดาวน์โหลดที่สูงกว่าผลจากแอปพลิเคชันอื่น ในขณะที่ค่าเฉลี่ยความเร็วในการอัปโหลดจาก 4 แอปพลิเคชันในบทความนี้มีค่าเฉลี่ยที่สูงกว่า เนื่องจากผลการศึกษานี้เป็นผลเฉลี่ยจากการทดสอบด้วย 4 แอปพลิเคชัน ซึ่งหนึ่งในนั้นคือแอปพลิเคชัน nPerf ซึ่งให้ค่าเฉลี่ยความเร็วในการอัปโหลดที่ค่อนข้างสูงกว่าผลการทดสอบจากแอปพลิเคชัน Speedtest

5.5 ในการศึกษานี้ คณะผู้วิจัยใช้เครื่องโทรศัพท์รุ่นที่เป็นเรือธง (Flagship) ของยี่ห้อหัวเหว่ย ที่ทำตลาดในช่วงกลางปี พ.ศ. 2563 ถึงต้นปี พ.ศ. 2564 ซึ่งมีประสิทธิภาพสูงเนื่องจากทำงานด้วยหน่วยประมวลผลกลางที่มีประสิทธิภาพสูง และใช้ซิมเลขหมายโทรศัพท์ของผู้ให้บริการ 2 ราย ที่เปิดบริการเป็นแพ็กเกจ 5G แบบไม่จำกัด จึงสามารถทำความเร็วในการทดสอบดาวน์โหลดและอัปโหลดข้อมูลได้ค่อนข้างสูงเมื่อเทียบกับค่าเฉลี่ยจากที่เคยมีการรายงาน (Fogg, 2021a; 2021b) อย่างไรก็ตาม ในความเป็นจริง ผู้ใช้งานบางส่วนอาจใช้เครื่องโทรศัพท์รุ่นที่มีประสิทธิภาพต่ำกว่าเครื่องที่ใช้ในการศึกษานี้ และผู้ให้บริการโครงข่ายโทรศัพท์เคลื่อนที่มักจะมีบริการหลายแบบหลายราคาให้ลูกค้าหรือผู้ใช้งานเลือก โดยบางแพ็กเกจ 5G อาจจำกัดความเร็วสูงสุดในการดาวน์โหลดข้อมูลไว้ที่ค่า ๆ หนึ่ง ดังนั้น หากมีการศึกษาเพิ่มเติมร่วมกับแพ็กเกจ 5G อื่นๆ ของผู้ให้บริการโครงข่าย 5G ครบทุกราย โดยให้ผู้ใช้งานทั่วไปมีส่วนร่วมในการทดสอบด้วยเครื่องโทรศัพท์เคลื่อนที่ของผู้ใช้งาน ก็มีความเป็นไปได้ค่อนข้างมากว่าความเร็วเฉลี่ยในการดาวน์โหลดและอัปโหลดข้อมูลในภาพรวมอาจมีค่าต่ำกว่าผลการศึกษาในบทความนี้

5.6 เมื่อนำค่าพารามิเตอร์คุณภาพของบริการที่วัดได้จากแต่ละแอปพลิเคชันมาเปรียบเทียบกันพบว่า ค่าพารามิเตอร์ที่ได้จากแต่ละแอปพลิเคชันมีค่าแตกต่างกัน และแม้ว่าค่าพารามิเตอร์คุณภาพของบริการที่ใช้ในการพิจารณาประสิทธิภาพของเครือข่ายจะมีเพียง 5 พารามิเตอร์หลัก แต่มีเพียงแอปพลิเคชัน Speedtest เพียงแอปพลิเคชันเดียวเท่านั้น ที่สามารถวัดและแสดงผลได้ทั้ง 5 พารามิเตอร์ (Speed Master แสดงค่าได้ครบทั้ง 5 พารามิเตอร์ แต่พบว่าแสดงค่าสูญเสียเป็นศูนย์ทุกครั้งที่ทดสอบ) สาเหตุของค่าที่มีความแตกต่างกัน อาจเกิดจากการมีวิธีการวัดที่แตกต่างกัน และมีเครื่องแม่ข่ายที่ใช้รองรับการทดสอบที่แตกต่างกันและอยู่ต่างสถานที่กัน

5.7 เมื่อพิจารณาถึงความน่าเชื่อถือของแต่ละแอปพลิเคชัน จากยอดดาวน์โหลดเพื่อติดตั้งใช้งานที่แสดงใน Play Store พบว่า Speedtest มีความน่าเชื่อถือมากที่สุด เนื่องจากมียอดดาวน์โหลดสูงที่สุด คือมากกว่า 100 ล้านครั้ง Speed Master และ Opensignal มียอดดาวน์โหลดที่ใกล้เคียงกันคือมากกว่า 10 ล้านครั้ง ส่วน nPerf มียอดดาวน์โหลดมากกว่า 1 ล้านครั้ง (Google Play, 2021a; 2021b; 2021c; 2021d)

5.8 สำหรับค่าเฉลี่ยเวลาแฝงและค่าเฉลี่ยจิตเตอร์ที่มีค่า 19 มิลลิวินาทีและ 8 มิลลิวินาที ถือว่าอยู่ในเกณฑ์ที่ค่อนข้างน้อย และเมื่อเทียบกับค่าเฉลี่ยเวลาแฝง 28.1 มิลลิวินาทีและค่าเฉลี่ยจิตเตอร์ 11.9 มิลลิวินาที ที่เคยมีการศึกษาในประเทศไทยในช่วงไตรมาสที่ 2 ของปี พ.ศ. 2563 ถือว่าดีขึ้นโดยเฉพาะอย่างยิ่งค่าเฉลี่ยเวลาแฝงที่ลดลงเกือบ 10 มิลลิวินาที (Daengsi & Wuttidittachotti, 2020)

5.9 ในกรณีของค่าร้อยละความสูญเสียจากการศึกษานี้ มีค่าเฉลี่ยรวมประมาณร้อยละ 0.3 ซึ่งเป็นค่าที่ค่อนข้างน้อย โดยเฉพาะอย่างยิ่งเมื่อเปรียบเทียบกับผลการศึกษาในประเทศไทยในช่วงไตรมาสที่ 2 ของปี พ.ศ. 2563 ที่มีค่าร้อยละความสูญเสียเท่ากับ 1.94 (Daengsi & Wuttidittachotti, 2020) ถือว่าลดลงจากเดิมมากกว่า 6 เท่า

5.10 สำหรับประเด็นเรื่องค่าเฉลี่ยจิตเตอร์และค่าเฉลี่ยร้อยละความสูญเสียที่มีค่าความเบี่ยงเบนมาตรฐานค่อนข้างสูงมาก สันนิษฐานว่ามีสาเหตุมาจากลักษณะการเกิดค่าจิตเตอร์และค่าความสูญเสียมีลักษณะการเกิดเป็น

ครั้งคราวหรือปะทุ (Burst) ขึ้นเป็นครั้งคราว ไม่ใช่การเกิดขึ้นเกิดขึ้นอยู่ตลอดเวลาหรือเกิดขึ้นอย่างต่อเนื่อง แต่เมื่อเกิดขึ้นแล้วอาจมีผลทำให้ค่าจิตเตอร์และค่าความสูญเสียในช่วงขณะนั้นมีค่าสูงกว่าเกณฑ์เฉลี่ยมาก

5.11. การศึกษานี้มุ่งเน้นศึกษาในมุมมองของผู้ใช้งานหรือผู้บริโภคที่ใช้บริการ 5G ด้วยเครื่องโทรศัพท์ 5G และเป็นการทดสอบในพื้นที่กรุงเทพมหานครและปริมณฑลที่ผู้ให้บริการโฆษณาประชาสัมพันธ์ว่าให้บริการ 5G อย่างไรก็ตาม จากผลการศึกษาที่แสดงในตารางที่ 3 จะเห็นได้ว่า ผลจากการทดสอบบริการ 5G ของ Oper1 และ Oper2 ประมาณ 200 จุด ตามสถานีรถไฟฟ้าบีทีเอสพบว่า สามารถใช้บริการ 5G ผ่านโครงข่ายที่เป็นระบบ 5G อยู่ที่ระดับร้อยละ 54.6 และ 60.8 ตามลำดับ ในขณะที่ร้อยละ 43.4 และ 39.2 เป็นการใช้บริการ 5G ผ่านโครงข่าย 4G ของ Oper1 และ Oper2 ตามลำดับ ซึ่งก็ถือว่าเป็นไปตามมาตรฐานของ ITU ที่กำหนดไว้ว่า เทคโนโลยี 5G จะต้องรองรับการทำงานร่วมกันกับโครงข่ายที่เป็นเทคโนโลยีเดิมได้ ไม่ว่าจะเป็น 4G หรือ 3G อย่างไรก็ตาม ในการวิเคราะห์และแสดงผลการทดสอบ จึงไม่มีการวิเคราะห์ผลแยกเป็น 5G กับ 4G เนื่องจาก ในมุมมองของผู้บริโภคที่ซื้อและใช้บริการ 5G ทั่วไป จะไม่ทราบหรือสนใจในรายละเอียดว่าบริการ 5G ที่ตนเองใช้งานอยู่นั้น จะเป็นโครงข่าย 5G จริงๆ หรือเป็นโครงข่าย 4G ที่ให้บริการแทน เนื่องจากจะเชื่อโดยปริยายว่าตนใช้งานผ่านโครงข่าย 5G ดังนั้นหน่วยงานกำกับดูแลอาจพิจารณาและศึกษาเพิ่มเติมว่าควรมีการกำหนดเกณฑ์เพิ่มเติมสำหรับตรวจสอบหรือประเมินคุณภาพของบริการ 5G ที่ให้บริการโดยผู้ให้บริการแต่ละรายในประเด็นนี้หรือไม่อย่างไร

5.12 ในประเด็นที่ไม่พบบริการ 5G SA ดังที่ได้แสดงในตารางที่ 3 ทั้งที่เครื่องโทรศัพท์ที่ใช้ในการทดสอบเป็นรุ่นที่รองรับ 5G SA (Khaosod Online, 2563) สันนิษฐานว่า สถานีฐานให้ที่ให้บริการตามแนวสถานีรถไฟฟ้าบีทีเอสบางสถานีฐานอาจยังไม่รองรับ 5G SA ในขณะที่บางสถานีฐานอาจจะรองรับ 5G SA แต่ไม่พร้อมให้บริการด้วย 5G SA ณ ขณะที่มีการทดสอบ อย่างไรก็ตาม ในประเด็นนี้ คณะผู้วิจัยได้รับทราบข้อมูลเพิ่มเติมในภายหลังจากสมาชิกรายหนึ่งของเพจกลุ่ม TelecomLover คนรักเสามือถือ 5G + เทสสปีด (ที่ทำการทดสอบคุณภาพสัญญาณ 4G/5G แล้วเผยแพร่บนเพจดังกล่าวเป็นประจำ ระบุว่า ผู้ให้บริการบางรายเปิดให้สามารถใช้งาน 5G SA ได้เฉพาะซิมของลูกค้าที่เปิดใช้งานกับเครื่องโทรศัพท์ที่เป็นยี่ห้อและรุ่นที่กำหนดเท่านั้น (กลุ่ม TelecomLoverฯ, 2021)

5.13 เมื่อพิจารณาในภาพรวมพบว่าในกรณีของการดาวน์โหลด แอปพลิเคชัน Speedtest วัดค่าเฉลี่ยความเร็วได้สูงที่สุด รองลงมาคือ nPerf และ 2 อันดับสุดท้ายคือ Speed Master และ Opensignal ซึ่งให้ผลการทดสอบใกล้เคียงกัน ส่วนกรณีของการอัปโหลดพบว่าแอปพลิเคชัน nPerf ให้ผลการทดสอบสูงกว่าแอปพลิเคชันอื่นอย่างเห็นได้ชัด โดยเฉพาะอย่างยิ่งเมื่อทดสอบกับ Oper1 ที่พบว่าผลการทดสอบที่ได้มีค่าสูงกว่าแอปพลิเคชันอื่นถึง 3 เท่า นอกจากนี้ ในการณีของค่าเวลาแฝงที่พบว่าแอปพลิเคชัน Speedtest ให้ค่าเฉลี่ยน้อยที่สุด รองลงมาคือ nPerf และ Opernsinal ซึ่งให้ผลการทดสอบใกล้เคียงกัน ในขณะที่ Speed Master ให้ผลการทดสอบ

พารามิเตอร์นี้แย่ที่สุด โดยมีค่าเฉลี่ยที่สูงกว่าแอปพลิเคชันอื่น 2-3 เท่า สำหรับกรณีของจิตเตอร์ที่พบว่าแอปพลิเคชัน Speedtest ทดสอบค่าจิตเตอร์ได้น้อยที่สุด (ดีที่สุด) น้อยกว่า Speed Master ประมาณ 2-3 เท่า และน้อยกว่า nPerf ประมาณ 3-5 เท่า ในขณะที่ Opensignal ไม่แสดงค่านี้ ซึ่งจากผลการศึกษาที่ได้จากแต่ละแอปพลิเคชันที่วัดค่าพารามิเตอร์ต่างๆ ได้แตกต่างกันนี้ สันนิษฐานว่าอาจเกิดจาก กระบวนการทดสอบที่มีรายละเอียดที่แตกต่างกัน เช่น ขนาดของแพ็กเกจและโพรโทคอลใช้ในการทดสอบ รวมไปถึงตำแหน่งที่ตั้งและประสิทธิภาพของเครื่องแม่ข่าย เป็นต้น ซึ่งปัจจัยเหล่านี้ล้วนมีผลต่อการทดสอบ

5.14 จากผลการศึกษานี้จะเห็นได้ว่าผลการทดสอบจากแต่ละแอปพลิเคชันมีความแตกต่างกันอย่างเห็นได้ชัด ดังนั้นจึงควรมีหน่วยงานกลางของรัฐ เช่น สำนักงานสำนักงานคณะกรรมการกิจการกระจายเสียง กิจการโทรทัศน์ และกิจการโทรคมนาคมแห่งชาติ (กสทช.) หรือมหาวิทยาลัยที่มีความน่าเชื่อถือ เป็นต้น ทำหน้าที่สำรวจในรูปแบบของคราวด์ซอร์สซิง (Crowdsourcing) เช่นเดียวกับผู้ให้บริการแอปพลิเคชันหลายรายแล้วจัดทำรายงานผลการศึกษาเพื่อรายงานต่อสาธารณชนอย่างสม่ำเสมอ ทั้งนี้เพื่อสร้างหรือเพิ่มความเชื่อมั่นให้กับผู้บริโภคตลอดจนนักลงทุนทั้งจากในประเทศและต่างประเทศ เนื่องจากระบบสื่อสารโทรคมนาคมเป็นโครงสร้างพื้นฐานที่สำคัญของประเทศ

5.15 การศึกษานี้ดำเนินการภายใต้ข้อจำกัดทั้งทางด้านงบประมาณ ทรัพยากร และเวลา จึงอาจกล่าวได้ว่า ผลการศึกษานี้เป็นเพียงผลการศึกษาเบื้องต้นเท่านั้น เพื่อเพิ่มความน่าเชื่อถือของผลการศึกษา จึงควรมีการศึกษาเพิ่มเติมด้วยเครื่องมือที่พร้อมและสอดคล้องหรือเทียบเคียงได้กับระเบียบวิธีวิจัยหรือวิธีดำเนินการทดสอบในการศึกษาที่เคยมีการดำเนินการก่อนหน้านี้ เพื่อให้สามารถนำผลการศึกษาที่ได้ ไปเปรียบเทียบกับผลการศึกษาเหล่านั้นได้ นอกจากนี้ ในอนาคตควรมีการศึกษาเพิ่มเติมเกี่ยวกับเครื่องมือมาตรฐานหรือวิธีการวัดที่เป็นมาตรฐานและเป็นที่ยอมรับและทำการวัดด้วยเครื่องมือหรือวิธีการดังกล่าว แล้วทำการศึกษาเปรียบเทียบกับผลการทดสอบที่ได้จากการวัดด้วย 4 แอปพลิเคชันที่ใช้ในการศึกษานี้ ทั้งนี้เพื่อพิสูจน์และเพิ่มความน่าเชื่อถือของผลการศึกษาในเชิงวิชาการ

## 6. สรุป

จากการศึกษาเกี่ยวกับคุณภาพของบริการของโครงข่าย 5G กรณีสถานีรถไฟฟ้าบีทีเอสในบทความนี้ จะเห็นได้ว่า เทคโนโลยี 5G ซึ่งจัดเป็นโครงสร้างพื้นฐานที่มีความสำคัญและมีความจำเป็นต่อเศรษฐกิจและสังคม ได้รับการพัฒนาขึ้นเป็นอย่างมาก โดยเฉพาะอย่างยิ่งในเขตพื้นที่กรุงเทพมหานครและปริมณฑล โดยมีค่าเฉลี่ยความเร็วในการดาวน์โหลดข้อมูลอยู่ที่ 240 เมกะบิตต่อวินาที และมีค่าความเฉลี่ยความเร็วในการอัปโหลดข้อมูลอยู่ที่ 87 เมกะบิตต่อวินาที โดยเฉพาะอย่างยิ่งเมื่อเทียบกับค่าเฉลี่ยที่เคยมีการศึกษาจากโครงข่าย 4G ในปี พ.ศ. 2563 พบว่ามีพัฒนาการแบบก้าวกระโดด และเมื่อเปรียบเทียบกับรายงานของ Opensignal ที่ทำการศึกษากับข้อมูลจากหลายประเทศในช่วงเวลาใกล้เคียงกัน พบว่า ประสิทธิภาพของโครงข่าย 5G ที่ได้จากการศึกษานี้มี

ประสิทธิภาพเป็นอันดับต้น ๆ ของภูมิภาคเอเชียแปซิฟิก อย่างไรก็ตามหากมีการศึกษาเพิ่มเติมในอนาคต อาจต้องมีการพิจารณาและทบทวนในประเด็นการเลือกใช้เครื่องมือ โดยเฉพาะอย่างยิ่งแอปพลิเคชันที่นำมาใช้ในการทดสอบเนื่องจากยังไม่ใช่เครื่องมือมาตรฐาน

**7. ข้อเสนอแนะ**

    **7.1 ข้อเสนอแนะสำหรับการศึกษาครั้งต่อไป**

        การศึกษานี้ยังจำกัดเฉพาะสถานีรถไฟฟ้าบีทีเอสซึ่งอยู่ในพื้นที่กรุงเทพมหานครและปริมณฑล ควรที่จะมีการศึกษาเพิ่มเติมในพื้นที่อื่นๆ ไม่ว่าจะเป็นพื้นที่อื่นในกรุงเทพมหานครเองหรือพื้นที่ต่างจังหวัด ทั้งนี้เพื่อให้ทราบถึงข้อเท็จจริงเกี่ยวกับการพัฒนาโครงข่าย 5G ของผู้ให้บริการในประเทศไทย และเป็นข้อมูลให้ผู้ให้บริการนำไปประกอบการพิจารณาในการพัฒนาโครงข่าย 5G ให้มีประสิทธิภาพมากขึ้นและครอบคลุมพื้นที่มากขึ้น นอกจากนี้หากต้องการเปรียบเทียบกับผลการศึกษาอื่นที่เคยมีการดำเนินการไปแล้ว เช่น ผลการศึกษาในต่างประเทศ ควรพิจารณาดำเนินการในรูปแบบเดียวกันหรือเทียบเคียงได้กับรูปแบบเดิม ทั้งนี้เพื่อเพิ่มความน่าเชื่อถือของผลการเปรียบเทียบ นอกจากนี้ ยังมีประเด็นเรื่องการเลือกใช้เครื่องมือในการศึกษาที่อาจมีการดำเนินในอนาคต ที่ควรเป็นเครื่องมือมาตรฐานที่ได้รับการยอมรับ

    **7.2 ข้อเสนอแนะเชิงนโยบายสำหรับกิจการสื่อสาร**

        จากผลการศึกษานี้จะเห็นได้ว่าผลการทดสอบจากแต่ละแอปพลิเคชันมีความแตกต่างกันอย่างเห็นได้ชัด ดังนั้นจึงควรมีหน่วยงานกลางของรัฐ เช่น สำนักงานคณะกรรมการกิจการกระจายเสียง กิจการโทรทัศน์ และกิจการโทรคมนาคมแห่งชาติ (กสทช.) หรือมหาวิทยาลัยที่มีความน่าเชื่อถือ ทำหน้าที่สำรวจในรูปแบบของคราวด์ซอร์สซิง (Crowdsourcing) เช่นเดียวกับผู้ให้บริการแอปพลิเคชันหลายราย แล้วจัดทำรายงานผลการศึกษาเพื่อเผยแพร่ต่อสาธารณะอย่างสม่ำเสมอ ทั้งนี้ เพื่อเป็นข้อมูลกลางสำหรับผู้ให้บริการโครงข่าย 5G แต่ละรายนำไปปรับปรุงและพัฒนาโครงข่ายของตนเองให้มีประสิทธิภาพดียิ่งขึ้น นอกจากนี้รายงานดังกล่าวยังช่วยเสริมสร้างหรือเพิ่มความเชื่อมั่นให้กับผู้บริโภค ตลอดจนนักลงทุนทั้งจากในประเทศและต่างประเทศได้ เนื่องจากปัจจุบันระบบสื่อสารโทรคมนาคมกลายเป็นโครงสร้างพื้นฐานที่สำคัญและเป็นหนึ่งในดัชนีชี้วัดความเจริญของประเทศ

**รายการเอกสารอ้างอิง**

**การขัดกันแห่งผลประโยชน์**

ผู้นิพนธ์ไม่มีผลประโยชน์ทับซ้อนทั้งด้านบวกและด้านลบ หรือมีความขัดแย้งกันระหว่างผลประโยชน์ส่วนตนและผู้มีส่วนได้ส่วนเสียจากการศึกษานี้